\documentclass[prb,footinbib,amsmath,amssymb,twocolumn,floatfix,showpacs]{revtex4}
\usepackage{graphicx}
\usepackage{graphics,color}
\usepackage{hyperref}

\newcommand{\bs}{\;\;\;\;\;}
\newcommand{\sms}{\;\;}
\newcommand{\ve}{\mathbf}

\newcommand{\D}{{\rm d}}

\begin{document}

\title{Tunable edge magnetism at graphene/graphane interfaces}
\author{Manuel J. Schmidt}
\author{Daniel Loss}
\affiliation{Department of Physics, University of Basel, Klingelbergstrasse 82, 4056 Basel, Switzerland}
\date{\today}
\pacs{73.22.Pr,75.75.-c,71.10.Pm,73.20.-r}


\begin{abstract}
We study the magnetic properties of graphene edges and graphene/graphane interfaces under the influence of electrostatic gates. For this, an effective low-energy theory for the edge states, which is derived from the Hubbard model of the honeycomb lattice, is used. We first study the edge state model in a mean-field approximation for the Hubbard Hamiltonian and show that it reproduces the results of the extended 2D lattice theory. Quantum fluctuations around the mean-field theory of the effective one-dimensional model are treated by means of the bosonization technique in order to check the stability of the mean-field solution. We find that edge magnetism at graphene/graphane interfaces can be switched on and off by means of electrostatic gates. We describe a quantum phase transition between an ordinary and a ferromagnetic Luttinger liquid - a realization of itinerant one-dimensional ferromagnetism. This mechanism may provide means to experimentally discriminate between edge magnetism or disorder as the reason for a transport gap in very clean graphene nanoribbons.
\end{abstract}

\maketitle

\section{Introduction}
Since graphene can routinely be isolated in the laboratory \cite{graphene1,graphene_rmp}, this monolayer of carbon atoms has received much attention because of its remarkable structural and electronic properties. One of the more recent graphene riddles is the one about the existence of edge magnetism. This phenomenon is based on a simple intuitive picture: zigzag edges of honeycomb lattices support so-called edge states, i.e. exponentially localized electronic states at the edges with nearly zero energy. The 'flatness' of the energy dispersion of the edge states leads to a high local density of states at the Fermi energy near the graphene edges. Therefore, these flat bands become susceptible to electron-electron interactions; the electronic system can lower its total energy, for instance, by polarizing the electron spin in the edge states. One consequence of edge magnetism is the appearance of a transport gap in narrow graphene nanoribbons (GNRs) which is inversely proportional to the GNR width. Indeed such a transport gap has been measured\cite{han_transport_gap}. However, it cannot be attributed unequivocally to edge magnetism. Other mechanisms, like Coulomb blockade in GNRs with rough edges\cite{coulomb_blockade}, have been shown to be also compatible with the experimental results.

Though a finite magnetization localized at edges of graphene nanoribbons (GNR) has never been directly measured, there seems to be a considerable consensus on the theory side about the presence of edge magnetism in clean GNRs: Hubbard model mean-field theories\cite{fujita_mf,jung_macdonald_mf,sasaki_saito_mf}, ab-initio calculations\cite{son_dft,pisani_dft}, and even theories that include quantum fluctuations\cite{feldner_mb,dutta_mb,hikihara_mb} consistently predict a ground state with a finite local magnetic moment at zigzag edges. Most of these calculations are numerical and are based on two-dimensional calculations on a honeycomb lattice. The actual one-dimensional character of edge magnetism has been appreciated only by a few authors (see Ref. \onlinecite{edge_magnetism_1D_1}).

In this paper, we show that the underlying mechanism of edge magnetism can be fully understood from a one-dimensional point of view, namely by an effective model which retains only the edge states while the bulk states are dropped. This is a considerable simplification but we will show that the deviation from a numerical two-dimensional lattice calculation is small. Our effective model is based on a simplified edge state model which we have used in an investigation of edge states at graphene/graphane interfaces\cite{our_EA_paper}. The crucial feature of this model is that it accounts for a finite bandwidth of the edge states. The advantages of this effective model are remarkable: (a) the mean-field theory is accessible analytically. (b) the impact of farther neighbor hoppings, electrostatic gates, graphene/graphane interfaces, which are encoded in the bandwidth of the effective edge state dispersion, on the edge magnetism can be investigated - also analytically. (c) quantum fluctuations can be included in a large parameter regime within the framework of bosonization.

In addition to this, our theory offers a parameter, the bandwidth of the edge states, by which the ferromagnetic transition can be driven. We propose a specific configuration of electrostatic gates at graphene/graphane interfaces which provides direct experimental access to this parameter so that the critical regime of this transition can directly be investigated experimentally. One of the big gains of such an experimental knob by which the edge magnetism can be turned on and off is that it provides a method to discriminate between different sources of transport gaps in clean graphene nanoribbons (terminated by graphane): if the transport gap is due to edge magnetism, it should disappear as the bandwidth of the edge states is increased sufficiently so that the edge magnetism is switched off. Disorder induced transport gaps, on the other hand, should be largely unaffected by this gate because it is designed not to change the carrier density in the graphene region.

This paper is organized as follows. In Section \ref{section_gate_dispersion} we discuss the effective edge state model and the appearance of a finite bandwidth on the basis of a single-particle picture. The various possible mechanisms which affect the edge state bandwidth are subsumed in a single effective parameter. Section \ref{section_hubbard} is dedicated to the mean-field treatment of the effective electron-electron interaction in the edge state model. In Section \ref{bosonization} we address the quantum fluctuations in the edge states by means of a bosonization technique. We close with a critical discussion of the applicability of our model and the experimental impact of our findings in Section \ref{section_discussion}.

\section{Gated edge state model\label{section_gate_dispersion}}
\begin{figure}[!ht]
\centering
\includegraphics[width=170pt]{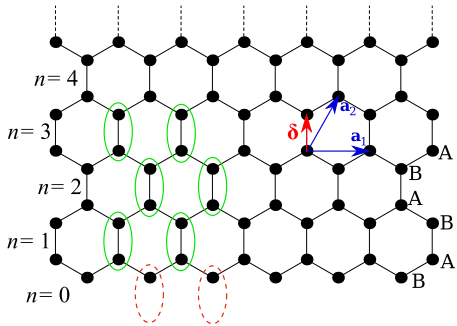}
\caption{(Color online) Lattice vectors $\ve a_1,\ve a_2$ and nearest neighbor vector $\boldsymbol\delta$ at an $\alpha$ edge. The coordinate $n\equiv n_2$ identifies the position perpendicular to the edge direction. The full (green) ellipses indicate the bulk unit cells ($n\geq 1$), while the dashed (red) circles indicate the truncated unit cell at the edge ($n=0$). The sublattice index $s=$A,B is also shown.}
\label{fig_definitions}
\end{figure}

In this section, we exemplarily consider an $\alpha$ edge in graphene or, equivalently, an $\alpha$ interface between graphene and graphane\cite{our_EA_paper}. For the $\beta$ edge\cite{our_EA_paper}, the findings are qualitatively similar, however some complications due to a possible commensurability with the lattice\footnote{Because the support of the $\alpha$ edge state in the Brillouin zone of a GNR is $k\in[\frac{2\pi}3,\frac{4\pi}3]$, the maximum momentum difference at the Fermi levels $k_{F,R} - k_{F,L}<\frac{2\pi}3$, so that first order umklapp processes are not allowed. For $\beta$ edge states the situation is different and for some filling fractions, umklapp scattering may well play an important role.} may arise. We choose a simplified description of graphene in which we only take into account nearest neighbor hopping between the $\pi$ orbitals
\begin{equation}
H_0 = t\sum_{\left<i,j\right>,\sigma} c_{i,\sigma}^\dagger c_{j,\sigma}
\end{equation}
with $t\simeq -3$eV. The operator $c_{i,\sigma}$ annihilates an electron with spin $\sigma$ at site $i$. Here, $i:=(n_1,n_2,s),\sms n_1,n_2\in\mathbb Z, s=0,1$ is a collective index which represents the lattice site at $\ve R_i = n_1\ve a_1+n_2\ve a_2+s\boldsymbol\delta$. $\ve a_1,\ve a_2$ are the Bravais lattice vectors and $\boldsymbol\delta$ is the vector which connects A ($s=0$) and B ($s=1$) sublattice sites (see Fig. \ref{fig_definitions}). $\left<i,j\right>$ runs over nearest neighbors on a honeycomb lattice. At the edges of the system, the sum must be restricted appropriately\cite{our_EA_paper}.

We aim at a description of systems which are lattice-translationally invariant along the $\ve a_1$ direction and it is convenient to transform this direction to k-space. Therefore, we work with the electron operators $d_{n,k,s,\sigma} = N_x^{-\frac12} \sum_{n_1} e^{-i k n_1} c_{n_1,n,s,\sigma}$, henceforth. $N_x$ is the number of unit cells in the $\ve a_1$ direction. It is well known\cite{fujita} that in this model, a zero-energy state exists which is exponentially localized at the $\alpha$ edge
\begin{equation}
\left|\psi_{k,\sigma}\right> =\mathcal N_k \left[ \sum_{n=0}^\infty e^{-n/\xi_k + i n \phi} d_{n,k,B,\sigma}^\dagger \right] \left|0\right>,
\end{equation}
where $\xi_k = -(\ln|u_k|)^{-1}$ is the localization length, $\mathcal N_k = (1-|u_k|^2)^{\frac12}$ is a normalization constant, and $u_k=1+e^{ik}$. $\phi$ is some unimportant phase. Note that this $\alpha$ edge state exists for $k\in\left[\frac{2\pi}3,\frac{4\pi}3\right]$, the rest of the Brillouin zone being the domain of the $\beta$ edge state\cite{our_EA_paper}.

We have shown in Ref. \onlinecite{our_EA_paper} that it is possible to incorporate some details of the edge states, like a finite bandwidth or spin-orbit interaction, in this simplified model on an effective level. The terms in the Hamiltonian which are responsible for these additional edge state properties are usually much smaller than $H_0$. Thus we may treat these small terms within perturbation theory by projecting them onto the subspace which is spanned by the edge states $\left|\psi_{k,\sigma}\right>$.

The dominant effect of a certain class of Hamiltonians describing, e.g. graphene/graphane interfaces\cite{our_EA_paper} or next-nearest neighbor hoppings\cite{distant_neighbor_hopping_dispersion}, is to create a non-zero bandwidth of the edge states. We model this class of bandwidth-generating edge properties as an effective local gate, described by the Hamiltonian
\begin{equation}
H_G  = -t_e \sum_{k,\sigma} d_{0,k,B,\sigma}^\dagger d_{0,k,B,\sigma}
\end{equation}
where $t_e$ is the amount by which the on-site energy of the outermost carbon atoms of the $\alpha$ edge are changed. Note that $t_e$ is an effective parameter which comprises, apart from the potential due to a real electrostatic gate, also many other details of the edge. As long as $|t_e|\ll |t|$, we may resort to first order perturbation theory in $H_G$ and find for the energy of the $\alpha$ edge state $\left|\psi_{k\sigma}\right>$
\begin{equation}
\epsilon_0(k) \simeq -t_e(2\cos(k-\pi) - 1)=-t_e \mathcal N_k^2,\bs k\in\left[\frac{2\pi}3,\frac{4\pi}3\right].\label{additional_dispersion}
\end{equation}
Usually an edge gate is not atomically sharp. However, as long as the gate is localized at the edge, the qualitative appearance of the dispersion is insensitive to the 'leakage' of the gate potential into the graphene bulk. 

A mechanism based on graphene/graphane interfaces which admits an especially well tunable dispersion of the form (\ref{additional_dispersion}) is discussed in Appendix \ref{appendix_graphane_gate}. The tunability of the bandwidth $t_e$ turns out to be crucial for changing the magnetic state of the edge. $t_e$ can be on the order of eV.

At this point, we would like to give some arguments why it is sufficient for the analysis of edge magnetism to keep only the edge states while the bulk states are completely removed from the considerations. These heuristic arguments are supplemented by a direct comparison between the effective model and a full lattice calculation in mean-field approximation (see Appendix \ref{appendix_numerics}).

\begin{figure}[!ht]
\centering
\includegraphics[width=220pt]{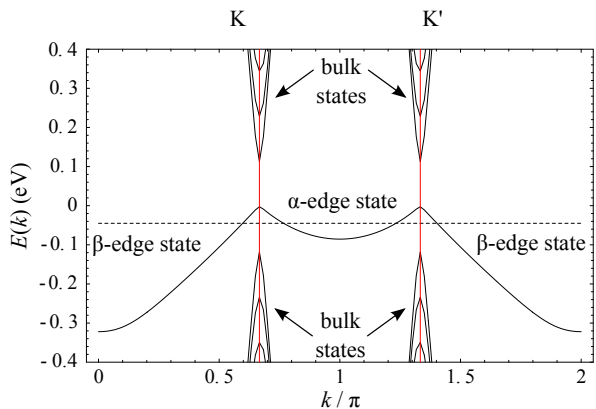}
\caption{(color online) Non-interacting band structure of a narrow $\alpha\beta$-ribbon (20nm wide) with graphane terminations. The calculation is based on a tight-binding model which takes into account the $\pi$- and $\sigma$-orbitals of the carbon sites\cite{our_EA_paper}. The dashed line indicates a typical Fermi energy (see text).}
\label{fig_ab_bandstructure}
\end{figure}

The most straightforward argument is based on the band structure of $\alpha\beta$ ribbons\cite{our_EA_paper}. At reasonable fillings (as shown in Fig. \ref{fig_ab_bandstructure}) only the edge states cross the Fermi energy while the bulk states are energetically remote. Since the $\alpha$ and $\beta$ edge states are localized at their respective edges, their mutual spatial overlap is exponentially small as long as $k$ does not get too close to one of the Dirac points K,K'. From the combination of the energy argument and the localization argument follows that the $\alpha$ edge state can be examined independently of all other states. The same argument holds also for the $\beta$ edge state.

How is the argument to be changed if, instead of $\alpha\beta$ ribbons, we consider the usual $\alpha\alpha$ ribbons? Then, instead of having a $\beta$ edge state attached to the $\alpha$ edge state at K and K', the edge state merges into a bulk state which is not exponentially localized at the other edge. However, the overlap between the edge state and this bulk state is still small (though not exponentially small) because the bulk state wave function is essentially proportional to $W^{-\frac12}\sin(\pi y/W)$, with $W$ the width of the ribbon, and thus is small where the edge state wave function is large.

Finally, one may ask what happens if we drop the restriction of narrow ribbons. In this case, the energy argument fails because it relied on the presence of a finite-size gap. Now, we resort to a rough scaling analysis of a density-density interaction like the Hubbard interaction, which is used below, or any screened interaction. In such a type of Hamiltonian, the spatial overlap of the wave function weights is important. The spatial density overlap of an edge state with localization length $\xi$ with the $n$th bulk state is roughly
\begin{equation}
o(n) \sim \frac1{W\xi} \int_0^\xi \D y \sin^2(\pi n y/W) \sim \frac{n^2\xi^2}{ W^3}.
\end{equation}
For increasing $W$, there are $\sim W$ bulk states at the Fermi energy so that the sum of all density overlaps of the edge state and the bulk states is $\sum_{n=1}^{\sim W} o(n) \sim W^0$, i.e. the total overlap does not grow with $W$. Thus, we expect that the bulk states do not become important as $W\rightarrow\infty$ because the spatial overlap between a typical edge state and the energetically relevant bulk states decreases sufficiently fast with $W$.

\section{Hubbard interaction in the projected model\label{section_hubbard}}
We want to study the electron-electron interaction effects in the edge states in the Hubbard approximation $H_U = U \sum_{i} \hat \rho_{i\uparrow} \hat \rho_{i\downarrow}$ with the local density operator $\hat\rho_{i\sigma} = c_{i\sigma}^\dagger c_{i\sigma}$. In terms of the $d_{n,k,s,\sigma}$ operators we have
\begin{equation}
H_U = U N_x^{-1}\sum_{n,q,s} \hat \rho_{n,q,s,\uparrow} \hat\rho_{n,-q,s,\downarrow},
\end{equation}
with $\rho_{n,q,s,\sigma} = \sum_{k} d_{n,k+q,s,\sigma}^\dagger d_{n,k,s,\sigma}$.

\subsection{Projection onto the effective edge state model}
Although the interaction energy scale $U$ can be quite large, we start with considering $H_U$ as a perturbation to $H_0$. We show in Appendix \ref{appendix_numerics} by a comparison to a less restricted numerical calculation that the following perturbative calculations capture the essential physics.

Since we aim at an effective description in terms of the edge states, disregarding the bulk states, we need to account for the background electron density, generated by a completely filled valence band. In the bulk, charge neutrality is reached if the complete valence band is filled and the conduction band is empty, i.e. one half of the $\pi$ band is filled. At half-filling there is on average one electron per site in the bulk system. In the bulk one can easily convince oneself that, because of the A B sublattice symmetry\footnote{We exclude interactions that are strong enough to create a bulk gap in graphene\cite{bulk_interaction_gap} thus breaking sublattice symmetry spontaneously. This may be done by placing the structure on a SiO$_2$ substrate.}, there really is exactly one electron per site - not only on average. At an edge, the sublattice symmetry is broken so that the site occupation may differ. Here, the edge states come into play. Half-filling is reached by occupying the valence band {\it and} one half of the edge states. The answer to the question, which half of the edge states have to be filled, depends on the energetic details of the edge states which are due to, e.g., spin-orbit interaction, dispersion and electron-electron interaction. If, for instance, we assume a non-magnetic configuration such as the one considered in Ref. \onlinecite{kane_mele} in which the spin-up states for $k\in[\frac{2\pi}3,\pi]$ and the spin-down states for $k\in[\pi,\frac{4\pi}3]$ are occupied, it turns out that also at the edges each site carries exactly one electron charge - one half with up-spin and one half with down-spin. The contribution to the total electron density per spin at unit cell $n$ and sublattice site $s$, coming from edge states that are occupied in this way is (note the invariance of the edge state wave function under $k\rightarrow 2\pi-k$)
\begin{equation}
\delta_{s,B}\int_{\frac{2\pi}3}^{\pi}\frac{\D k}{2\pi} \mathcal N_k^2 \left[2 (1+\cos k)\right]^n =: \frac12 \rho_0(n,s)\label{half_filled_edge_density},
\end{equation}
where $\rho_0(n,s)$ is the electron density per spin due to a fully filled edge state band. The background electron density $\rho_B(n,s,\sigma)$ (relative to half-filling) of a filled valence band and an empty edge state band is therefore given by
\begin{equation}
\rho_B(n,s,\sigma) = -\frac12 \rho_0(n,s). \label{background_density}
\end{equation}
A more rigorous derivation of Eq. (\ref{background_density}) can be found in Appendix \ref{appendix_background_density}.

In the remainder of this work, we will drop the sublattice index $s$, setting it to $B$. This is because the $A$ sublattice sites are always half-filled in the perturbative treatment and thus do not contribute to the following considerations.

\subsection{Polarized vs. unpolarized state}

In order to gain a rough overview over the basic principles from which the magnetic properties of the graphene edges derive, we start with comparing the energy of a ground state configuration with fully polarized edge states $\left|P\right> = \prod_k e^\dagger_{k,\uparrow} \left|\Omega\right>$ to the energy of an unpolarized configuration $\left|U\right> = \prod_{\sigma} \prod_{\frac{5\pi}6\leq k\leq \frac{7\pi}6} e_{k,\sigma}^\dagger\left|\Omega\right>$. $\left|\Omega\right>$ represents the completely filled valence band, which plays the role of the vacuum in the effective model. The operator $e_{k,\sigma}^\dagger$ creates one edge state with crystal momentum $k$ along the edge and spin $\sigma$.

The total kinetic energies (per unit length) of these states, relative to the kinetic energy of a completely filled valence band, are
\begin{eqnarray}
t_e E_P &:=& \left<P|H_{0,G}|P\right> = -t_e \int_{\frac{2\pi}3}^{\frac{4\pi}3} \frac{\D k}{2\pi}\mathcal N_k^2\\
t_e E_U &:=& \left<U|H_{0,G}|U\right> = -2 t_e  \int_{\frac{5\pi}6}^{\frac{7\pi}6} \frac{\D k}{2\pi} \mathcal N_k^2,
\end{eqnarray}
where the numerical factors $E_P$ and $E_{U}$ have been defined for later convenience. The Hamiltonian of the kinetic energy $H_{0,G} = H_0+H_G$ consists of the usual hopping Hamiltonian of graphene $H_0$ and of the gate Hamiltonian $H_G$, which is responsible for the finite bandwidth of the edge states.

For the calculation of the Hubbard energy, we treat all electron densities relative to half-filling. In the fully polarized state $\left|P\right>$ the spin-dependent electron density is given by
\begin{equation}
\rho_P (n,\sigma) = \sigma \frac12 \rho_0(n).
\end{equation}
In the unpolarized state $\left|U\right>$, the electron density is spin-independent

\begin{multline}
\rho_U(n,\sigma) = 
\int_{\frac{5\pi}6}^{\frac{7\pi}6} \frac{\D k}{2\pi} \mathcal N_k^2\left[2(1+\cos k)\right]^n - \frac12 \rho_0(n)\\ = \int_\pi^{\frac{4\pi}3} \frac{\D k}{2\pi} \underbrace{{\rm sign}\left(\frac{7\pi}6-k\right)}_{=:s_k}\mathcal N_k^2\left[2(1+\cos k)\right]^n.
\end{multline}
From the electron densities, the interaction energies are given by $U\sum_{n=0}^\infty \rho_X(n,\uparrow)\rho_X(n,\downarrow)$, where $X=P,U$. This leads to
\begin{equation}
U W_P := -U \int_{\pi}^{\frac{4\pi}3} \frac{\D k\D k'}{(2\pi)^2} \frac{(\mathcal N_k \mathcal N_{k'})^2}{1-4(1+\cos k)(1+\cos k')}
\end{equation}
for the interaction energy of the polarized state and
\begin{equation}
U W_U := U \int_\pi^{\frac{4\pi}3} \frac{\D k\D k'}{(2\pi)^2} \frac{s_k s_{k'}(\mathcal N_k \mathcal N_{k'})^2}{1-4(1+\cos k)(1+\cos k')}
\end{equation}
for the interaction of the unpolarized state.

Obviously, for flat bands ($t_e=0$), the polarized state $\left|P\right>$ is the ground state. If the bandwidth $t_e$ is increased to positive values, however, the total energy of the polarized state $E_Pt_e + W_P U$ eventually becomes larger than the total energy of the unpolarized state. This happens at
\begin{equation}
\left[\frac{t_e}U\right]_{\rm crit.} \sim \frac{W_{P} - W_U}{E_P - E_{U}} \simeq 0.17\label{critical_alpha_edge_dispersion},
\end{equation}
the critical bandwidth/interaction ratio for positive $t_e$ at which the system becomes unpolarized. For negative $t_e$, the polarized state becomes unstable with respect to another unpolarized state that has the inverse edge state occupation of $\left|U\right>$. This instability occurs at the symmetric position $t_e/U \simeq - 0.17$.

\subsection{Solution of the mean-field equations}
The argumentation in the previous paragraph is of course very superficial since we have only compared the total energies of completely polarized and completely unpolarized states, neglecting the possibility of partial polarization. Thus, we now formulate a closed set of mean-field equations for the effective edge state model with Hubbard interaction and solve them.

At zero temperature, the mean-field energy of an edge state $\left|\psi_{k\sigma}\right>$, with all other edge states with mean-field energies smaller than $\epsilon_F$ occupied, is given by
\begin{multline}
\epsilon_\sigma(k) = \epsilon_0(k) \\+ U \int_{\frac{2\pi}3}^{\frac{4\pi}3} \frac{\D k'}{2\pi} \Gamma(k,k',0) \left[\Theta(\epsilon_F - \epsilon_{-\sigma}(k'))-\frac12\right], \label{sce1}
\end{multline}
where we have defined the effective interaction vertex
\begin{equation}
\Gamma(k,k',q) =  \frac{\mathcal N^\alpha_{k+q} \mathcal N^\alpha_{k} \mathcal N^\alpha_{k'-q} \mathcal N^\alpha_{k'}}{1-u_{k+q}^* u_k u^*_{k'-q} u_{k'}}.
\end{equation}
$\Gamma(k,k',0)$ results from the summation of the probability densities of two edge states, $k$ and $k'$, over the sites in $\ve a_2$ direction. For later convenience, $\Gamma$ has been introduced in a more general form than actually needed here, namely for non-zero $q$. It is important to note the invariance of $\Gamma(k,k',0)$ under $k\rightarrow 2\pi-k$ and $k'\rightarrow 2\pi-k'$, as well as under $k\leftrightarrow k'$.

\begin{figure}[!ht]
\centering
\includegraphics[width=210pt]{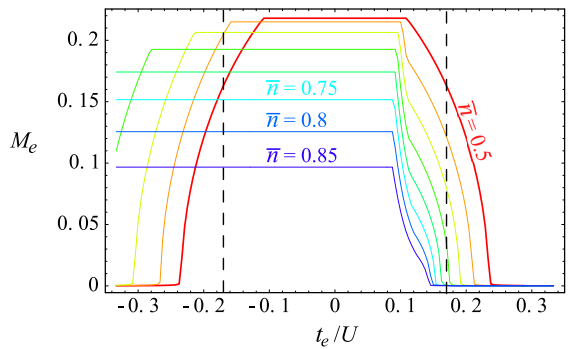}
\caption{(Color online) Mean-field result for the edge magnetization $M_e$ as a function of the bandwidth $t_e$ of the edge state. The topmost, thick (red) line represents the result for half-filling. The other lines show the results for increasing the filling from 1/2 (topmost line, red) to 0.85 (bottom line, blue). The dashed vertical lines indicate the critical bandwidth obtained from the energy comparison of polarized and unpolarized states at half-filling.}
\label{fig_mean_field_magnetization}
\end{figure}

The Fermi energy is fixed by a constant effective electron density $\bar n$. $\bar n=0$ means that the edge states are completely unoccupied and $\bar n=1$ means that all edge states are occupied.
\begin{equation}
\frac{2\bar n}3 = \sum_{\sigma} \int_{\frac{2\pi}3}^{\frac{4\pi}3} \frac{\D k}{2\pi} \Theta(\epsilon_F - \epsilon_\sigma(k))\label{sce3}
\end{equation}
Eqs. (\ref{sce1}) and (\ref{sce3}) are a complete set of mean-field equations which can easily be solved. It is convenient to characterize the solution by the edge magnetization which we introduce as
\begin{equation}
M_e = \sum_\sigma \sigma \rho(0,\sigma),\label{edge_magnetization}
\end{equation}
where $\rho(n,\sigma)=\int\frac{\D k}{2\pi} \Theta(\epsilon_F-\epsilon_\sigma(k)) \left|\left<0\right|d_{n,k,B,\sigma}\left|\psi_{k,\sigma}\right>\right|^2$ is the spin-dependent density at site $n$.

Fig. \ref{fig_mean_field_magnetization} shows the solutions of the mean-field equations for different fillings $\bar n$. One observes a second order quantum phase transition at a critical $|t_e/U|_{\rm crit.}$, below which the edge becomes spontaneously polarized. The polarization saturates as $|t_e/U|\rightarrow0$. For half-filling, the saturated regime corresponds to the usual edge magnetism\cite{fujita_mf, jung_macdonald_mf, sasaki_saito_mf, son_dft, pisani_dft} which is characterized by a complete polarization of the edge state spin. If $\bar n>\frac12$, also the minority spin edge state becomes partially occupied until, for $\bar n=1$, both spin species are completely filled and no spin-polarization exists any longer.

Note that, in addition to the phase transition at which the polarization starts to grow from zero, there is also another transition at which the polarization starts to deviate from the saturation value. While the first transition is of a Stoner type, the latter is not. As an illustration of the mean-field solution we provide three movies\cite{movies} showing the single-particle energy $\epsilon_\sigma(k)$ (see Eq. (\ref{sce1})) for both spins together with the edge magnetization for three different filling factors $\bar n = \frac14,\frac12,\frac34$. It can be seen that the situations $\bar n=\frac14,\frac34$ are connected by particle-hole symmetry. Furthermore, we see that, at the Stoner transition, the Fermi points of the two spin species start to become more and more split. At the second transition, which is not of Stoner type, the number of Fermi points changes from two to four.

In the following, we distinguish between the saturated regime and the regime close to $\left[t_e/U\right]_{\rm crit.}$ where the spin-polarization is small. The latter is called the regime of weak ferromagnetism. This regime will be especially important for the analysis of quantum fluctuations. Note that the stability of weak ferromagnetism on the mean-field level is a consequence of the special momentum-dependent form $\Gamma(k,k',q)$ of the effective edge state model. For a truly one-dimensional Hubbard model, the interaction vertex would be constant in $k$, so that a solution of the self-consistency equations does not yield weak ferromagnetism. We note that the mechanism we describe here is different from the one Bartosch {\it et al.} use \cite{bartosch} for the stabilization of weak ferromagnetism in one-dimensional metals.

\subsection{Weak ferromagnetic regime}
The general mean-field equations have been solved numerically in the previous subsection. Close to the critical point, however, one can obtain an approximate analytical solution with
\begin{equation}
\tau=\left[\frac{t_e}U\right]_{\rm crit.}-\frac{t_e}U
\end{equation}
as a small parameter. This will be useful below where quantum fluctuations around the mean-field solution in the weak ferromagnetic regime (but not too close to the transition) shall be analyzed by means of the bosonization technique.

\begin{figure}[!ht]
\centering
\includegraphics[width=220pt]{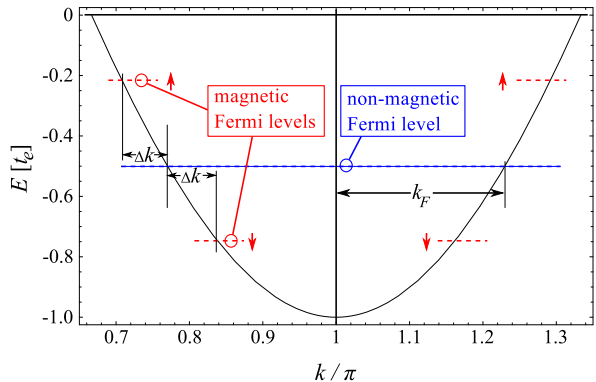}
\caption{Fermi levels in the non-magnetic ground state and in the magnetic ground state where the Fermi momenta $k_{Fr\sigma}$ are split by 2$\Delta k$ in k-direction.}
\label{fig_edge_state_fermi_levels}
\end{figure}

Weak ferromagnetism is characterized by a small imbalance in the spin population, which is quantified by spin-dependent Fermi momenta $k_{Fr\sigma} =\pi + r (k_F + \sigma \Delta k)$ (see Fig. \ref{fig_edge_state_fermi_levels}). Here, $r=R\, (L)$ stands for right-moving (left-moving) parts of the edge state dispersion at the Fermi energy. Henceforth, $R=+1$ and $L=-1$ when used in equations. Likewise, $\sigma=\pm1$ stands for up- and down-spin, respectively. We use the symbol $k_F$ for the distance of the two Fermi points from $\pi$ in the non-magnetic configuration, where $\Delta k=0$. A fixed $k_F$ corresponds to a fixed edge state filling $\bar n$, in both, the magnetic and the non-magnetic ground state.

For non-zero $\Delta k$, the kinetic energy and interaction energy are
\begin{eqnarray}
E_{\rm kin} &=& \int_{-\Delta k}^{\Delta k} \frac{\D k}{\pi} {\rm sign}(k) \bar\epsilon(\pi+k_F+k)\label{e_kin}\\
E_{\rm int} &=& U\int_{k_{FL\uparrow}}^{k_{FR\uparrow}} \frac{\D k_{\uparrow}}{2\pi}\int_{k_{FL\downarrow}}^{k_{FR\downarrow}} \frac{\D k_{\downarrow}}{2\pi}\Gamma(k_\uparrow,k_\downarrow,0)\label{e_int},
\end{eqnarray}
where we have defined the single particle kinetic energy, corrected by the background charge density of the filled valence band, $\bar\epsilon(k) = \epsilon_0(k) + U n_B(k)$, with
\begin{equation}
n_B(k) = - \int_\pi^{\frac{4\pi}3}\frac{\D k'}{2\pi} \Gamma(k,k',0).
\end{equation}
We approximate for small $\Delta k$
\begin{eqnarray}
\frac{\D E_{\rm kin}}{\D \Delta k} &\simeq& (t_e\alpha + U\beta )\Delta k\label{d_ekin_d_k} \\
\frac{\D E_{\rm int}}{\D \Delta k} &\simeq& -U (\gamma_1\Delta k - \frac16 \gamma_3\Delta k^3)
\end{eqnarray}
where we have dropped terms of order $\Delta k^3$ in the kinetic energy and terms of order $\Delta k^5$ in the interaction energy. The $O(\Delta k^3)$ terms in the kinetic energy could have been taken into account but it turns out that they only lead to an inessential quantitative renormalization of $\Delta k$. This approximation is equivalent to linearizing the single particle dispersion $\bar\epsilon(k)\simeq v_F (k-k_{Fr\sigma})$ around the Fermi points.

In the ferromagnetic regime, the total energy assumes a minimum for
\begin{equation}
\Delta k \simeq \pm \left[\frac{6\alpha}{\gamma_3} \tau\right]^{\frac12},\bs\text{for }\tau>0.\label{mfsol}
\end{equation}
If we had taken the third order terms in Eq. (\ref{d_ekin_d_k}) into account ($\frac{\beta_3 U}6+\frac{\alpha_3 t_e}6$), the right hand side of Eq. (\ref{mfsol}) would have been multiplied by a factor of $\sim$1.02. Thus, it is reasonable to neglect the curvature of $\bar\epsilon(k)$.

\begin{figure}
\centering
\includegraphics[width=200pt]{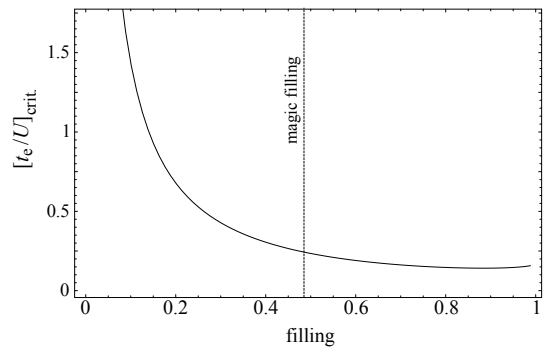}
\caption{Critical bandwidth $[t_e/U]_{\rm crit.}$ as a function of the edge state filling. 0 stands for a completely empty edge band ($k_{Fr}=\pi$ in the non-magnetic configuration) and 1 for a completely filled edge band ($k_{Fr}=\pi+r\pi/3$ in the non-magnetic configuration).}
\label{fig_tOverU_crit}
\end{figure}

The parameters $\alpha,\beta,\gamma_1,\gamma_3$ can be calculated from Eqs. (\ref{e_kin}) and (\ref{e_int}). This is especially convenient for a 'magic' filling, corresponding to $k_F=\arccos\frac78$, which is slightly lower than half-filling. Throughout the remainder of this paper, all analytic discussions are based on this filling. The reason for this is simply that the formulas are only about one third as long as for half-filling, for instance. We find
\begin{eqnarray}
\alpha  &=& \frac{\sqrt{15}}{2\pi} \simeq 0.616\\
\beta &=& \frac{\sqrt5}{3\pi^2}(4\sqrt3 \pi - 21) \simeq 0.0578\\
\gamma_1 &=& \frac{4}{5\pi^2} \left(10\sqrt 15 \arccos\frac78 - 17\right) \simeq 0.209 \\
\gamma_3 &=& \frac{660826-329000 \sqrt{15}\arccos\frac78}{875\pi^2} \simeq 1.96
\end{eqnarray}
and thus a critical bandwidth
\begin{multline}
\left[\frac{t_e}U\right]_{\rm crit.} = \frac{\gamma_1-\beta}\alpha \\= 2\frac{\sqrt3(175-68\sqrt5)+200\pi-600\arcsin\frac78}{75\pi} \simeq 0.244.
\end{multline}
We note that these parameters can be calculated analytically also for general fillings. However, the results are quite cumbersome. We therefore Fig. \ref{fig_tOverU_crit} shows the critical bandwidth as a function of the edge band filling.

Note that we describe essentially a Stoner mechanism here. This is why we will call the critical point $\tau=0$ the Stoner point, henceforth.

\section{Quantum fluctuations\label{bosonization}}
In the previous section, we have analyzed the mean-field theory of the effective edge state model. Now, we proceed by analyzing the stability of the mean-field solution with respect to quantum fluctuations. Since the effective model is one-dimensional, quantum fluctuations may be treated most easily by means of the bosonization technique\cite{giamarchi}. We have shown in the preceding section that the linearization of the single-particle dispersion around the Fermi points is a good approximation. This is an essential prerequisite for the applicability of bosonization.

With the annihilation operator $e_{k,\sigma}$ of an edge state with momentum $k$ along the edge and spin $\sigma$, the full Hamiltonian of the effective edge state model can be written as
\begin{multline}
H = \sum_{k,\sigma}'\left[\epsilon_0(k) + U n_B(k)\right] e_{k\sigma}^\dagger e_{k\sigma} \\ + \frac{U}{N_x}\sum_{k,k',q}' \Gamma(k,k',q) e^\dagger_{k+q\uparrow} e_{k\uparrow}e_{k'-q\downarrow}^\dagger e_{k'\downarrow},
\end{multline}
where the primed sum means that the summation is restricted such that $\frac{2\pi}3 \leq k,k',k+q,k'-q\leq \frac{4\pi}3$. For the derivation of the interaction term, see Appendix \ref{appendix_eff_hubbard}. The background term $n_B(k)$ describes the additional dispersion coming from the electron density of the filled valence band. We introduce a normal ordering
\begin{equation}
:A: \,\equiv  A - \left<\phi_0|A|\phi_0\right> ,
\end{equation}
where $\left|\phi_0\right>$ denotes the Slater-determinant of the ground state of the mean-field approximation to $H$, as discussed in the preceding section. With the generalized density 
\begin{equation}
n_\sigma(k) = \sum_{k'} \Gamma(k,k',0)\left<\phi_0\right|e_{k'\sigma}^\dagger e_{k'\sigma}\left|\phi_0\right>,
\end{equation}
the mean-field part of $H$ becomes
\begin{equation}
H_0 = \sum_{k,\sigma}\underbrace{\left[\epsilon_0(k) + U (n_B(k)+n_{-\sigma}(k))\right]}_{=:\epsilon_\sigma(k)} e^\dagger_{k\sigma} e_{k\sigma}\label{eq_mean_field_bos}
\end{equation}
and the term describing the quantum fluctuations around the solution of $H_0$ reads
\begin{equation}
H_1 = \frac U{N_x}\sum_{k,k',q}' \Gamma(k,k',q): e_{k+q\uparrow}^\dagger e_{k\uparrow} : : e_{k'-q\downarrow}^\dagger e_{k'\downarrow} :.
\end{equation}
Note that the mean-field energy $\epsilon_\sigma(k)$, defined in Eq. (\ref{eq_mean_field_bos}) is consistent with Eq. (\ref{sce1}).

For the bosonization we need to distinguish between the non-magnetic mean-field phase, which will turn out to behave as an ordinary spinful Luttinger liquid, and the magnetic phase, the properties of which are somewhat more intriguing. Especially the boundary of these phases will turn out to be complicated so that, in this work, we restrict the discussion to values of $t_e/U$, sufficiently far from the critical point. However, $t_e/U$ is still required to be larger than the value at which the spin-polarization of the edge-states saturates.

\subsection{Non-magnetic phase}

In the non-magnetic phase, i.e. for $t_e/U  > \left[t_e/U\right]_{\rm crit.}$, both spin species are equally occupied so that $\epsilon_\sigma(k)$ is spin-independent. The Fermi velocity is given by
\begin{multline}
v_F = \frac{\D}{\D k} \epsilon_\sigma(k) =2t_e \sin(k_F) \\+ U\int_{-\frac\pi3}^{\frac\pi3} \frac{\D k}{2\pi} \left[\Theta(k_F - |k|) - \frac12\right] \frac{\D}{\D k_F}\Gamma(\pi+k_F,\pi+k,0).
\end{multline}
An essential simplification needed in order to express $H_0$ by means of bosonic degrees of freedom is the linearization of the edge state dispersion $\epsilon_\sigma(k)$ around $\epsilon_F$ at the left-moving ($r=L$) and right-moving ($r=R$) Fermi point, i.e.
\begin{equation}
\epsilon_\sigma(k) \simeq v_F r (k-k_{Fr\sigma}),\sms\text{ for } |k-k_{Fr\sigma}|\ll k_F.
\end{equation}

\begin{figure}[!h]
\centering
\includegraphics[width=180pt]{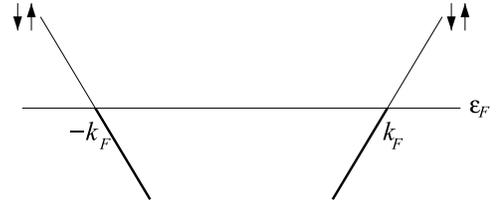}
\caption{Mean-field configuration of the non-magnetic regime. Both spin species are equally occupied (indicated by the bold black lines).}
\end{figure}

At the magic filling, to which we want to restrict the discussion, the Fermi velocity is
\begin{equation}
v_F = t_e \frac{\sqrt{15}}4  - U \delta
\end{equation}
with
\begin{equation}
\delta = \frac{\sqrt{15}}4 \left[\frac{t_e}{U}\right]_{\rm crit.}-\frac{3}{5\pi}\simeq 0.046
\end{equation}

Because in the Hubbard model only densities of different spin projections interact with each other one finds that\footnote{We use the notation from Ref. \onlinecite{giamarchi}.} $g_{i\parallel}=0,\sms\forall i$. For $g_{i\perp}$, the g-ology for the Hubbard interaction in the edge state model (see Appendix \ref{appendix_eff_hubbard}) gives
\begin{equation}
 g_{1\perp}=g_{2\perp}=g_{4\perp} = U \Gamma(\pi+k_F,\pi+k_F,0) = \frac{3U}5.
\end{equation}
Following the standard procedure of Abelian bosonization (see, e.g., Ref. \onlinecite{giamarchi}), one finds
\begin{equation}
H = H_c^0 + H_s^0 + H_s^1
\end{equation}
where the free Hamiltonians $H_c^0$ and $H_s^0$ for the charge and spin sector, respectively, are
\begin{equation}
H_\nu^0 = \frac {1}{2\pi} \int\D x \left[u_\nu K_\nu(\partial_x\theta_\nu(x))^2 + \frac{u_\nu}{K_\nu}(\partial_x \phi_\nu(x))^2\right],\label{bosonization_non_mag_h0}
\end{equation}
where $\nu=c,s$ with
\begin{equation}
u_c K_c = v_F\bs\text{and}\bs \frac{u_c}{K_c}=1+\frac{3U}{5\pi v_F}
\end{equation}
for the charge sector and
\begin{equation}
u_s K_s = v_F\bs\text{and}\bs \frac{u_s}{K_s}=1-\frac{3U}{5\pi v_F}
\end{equation}
for the spin sector. The bosonic fields $\phi_\nu(x)$ and $\theta_\nu(x)$ satisfy the commutation rules
\begin{equation}
\left[\phi_\nu(x),\partial_x\theta_{\nu'}(x')\right] = i\pi \delta_{\nu\nu'}\delta(x-x').
\end{equation}
The backscattering process $g_{1\perp}$ leads to a sine-Gordon term in the spin sector
\begin{equation}
H_s^1 = \frac{2g_{1\perp}}{(2\pi\eta)^2} \int \D x \cos(2\sqrt2 \phi_s(x)),\label{bosonization_non_mag_h1}
\end{equation}
where $\eta\sim k_F^{-1}$ is an ultraviolet cutoff\cite{giamarchi}.

Due to spin-charge separation, the spin and charge sectors may be treated separately. We want to focus on the spin sector here. The spin velocity $u_s$ becomes singular as $\frac{3U}{5\pi v_F}\rightarrow1$. This is exactly the Stoner point of the mean-field theory. Note that the Stoner point is also reflected in the spin susceptibility of the free bosonic theory, i.e. the theory in which the backscattering term $H_s^1$ is disregarded, $\chi_0 = \frac{K_s}{2\pi u_s} \propto \frac{1}{|5 \pi v_F - 3U|}.$
$\chi_0$ has a singularity at $\tau=0$.

It is well known\cite{giamarchi} that a perturbative treatment of $H_s^1$ leads to a renormalization of $K_s$ and the amplitude of the backscattering process $g_{1\perp}$. The $g_{1\perp}$ process is marginally irrelevant for SU(2) invariant systems like the edge state in the non-magnetic regime. This means that $K_s\rightarrow K_s^* = 1$ and $g_{1\perp}\rightarrow0$, as the ultraviolet part of the Brillouin zone is integrated out successively. In the renormalized theory, the spin susceptibility becomes
\begin{equation}
\chi_0 = \frac{K_s^*}{2\pi u_s} \propto \frac{1}{\sqrt{|5 \pi v_F - 3U|}}.
\end{equation}
However, the renormalization group for the backscattering process is perturbative in $g_{1\perp}/u_s$ and cannot be used close to the Stoner point, where $u_s\rightarrow0$. This is why we exclude the critical region from our argumentation in this work.

\subsection{Weak ferromagnetic regime}
In the previous subsection, we found that the energy which has to be paid for fluctuations of the field $\phi_s(x)$ is proportional to $(1-3U/5 \pi v_F)(\partial_x \phi_\sigma(x))^2$. The sign of this term becomes negative beyond the Stoner point. This means that the system can lower its energy by developing spin fluctuations and becomes unstable against a new ground state with a spontaneously broken symmetry. However, the bosonic theory based on Eqs. (\ref{bosonization_non_mag_h0}) - (\ref{bosonization_non_mag_h1}) is not able to actually predict the proper ground state. Therefore, we need to go back to the level of the mean-field theory. 

We have shown in the previous section that the mean-field theory becomes unstable with respect to a spin-polarized ground state at the Stoner point. Thus, it is reasonable for $t_e/U < \left[t_e/U\right]_{\rm crit.}$ to bosonize the quantum fluctuations around the ferromagnetic mean-field theory, rather than the non-magnetic. The question to be answered is then, if this polarized mean-field theory is stable with respect to quantum fluctuations.

\begin{figure}[!h]
\centering
\includegraphics[width=180pt]{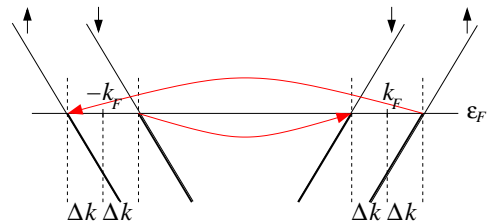}
\caption{Mean-field configuration in the weak ferromagnetic regime. The fat lines represent the occupied states and the thin lines represent the unoccupied states. The red arrows indicate the backscattering process, which is not momentum-conserving here.}
\label{fig_mag_kf}
\end{figure}

We restrict the discussion to the regime of weak ferromagnetism, i.e. we exclude the saturated regime. It should be noted that the different spin species may have slightly different Fermi velocities. This leads to a term $(v_{F\uparrow} - v_{F\downarrow})(\theta_s'\theta_c'+\phi_s'\phi_c')$ in the bosonized Hamiltonian. This term mixes the spin sector and the charge sector. However, $v_{F\uparrow}-v_{F\downarrow}$ is of the order of $\Delta k$ which is small in the weak ferromagnetic regime. Thus, we neglect this difference in the remainder of this work, keeping only the mean-value $v_F = \frac12(v_{F\uparrow}+v_{F\downarrow})$. We begin with the amplitudes of the forward scattering processes which lead to the quadratic part of the bosonic Hamiltonian. In the non-magnetic mean-field ground state, these two amplitudes are $g_{2\perp}=g_{4\perp}=U\Gamma(\pi+k_F,\pi+k_F,0)$. In the ferromagnetic mean-field ground state, characterized by a finite $\Delta k$, however, the amplitudes are
\begin{multline}
g_{2\perp}=g_{4\perp} =: U\Gamma_0(\Delta k) \\= U\Gamma(\pi+k_F+\Delta k,\pi+k_F-\Delta k,0)
\end{multline}
$\Gamma_0(\Delta k)$ has a maximum at $\Delta k=0$, which means that the interactions between the electrons at the Fermi level become weaker as the polarization $\sim\Delta k$ becomes larger. On the mean-field level this leads to the balance between the kinetic and the interaction energy which allows the existence of the weak ferromagnetic regime. In the framework of bosonization, the polarization dependence of the interaction has the important function of restricting $u_s/K_s$ to positive values: for small $\Delta k$ we find (for the magic filling)
\begin{equation}
\Gamma_0(\Delta k) \simeq \frac35 - \frac{136}{50}\Delta k^2
\end{equation}
so that for $\tau>0$
\begin{equation}
\frac{u_s/v_F}{K_s} \simeq \left(\frac{136 \alpha}{5\gamma_3} - \frac{5\pi}4\sqrt{\frac53}\right)\tau \simeq + 3.5\tau.
\end{equation}
Thus, the free bosonic theory in the ferromagnetic regime is meaningful if the proper mean-field theory is used.

\begin{figure}[!ht]
\centering
\includegraphics[width=220pt]{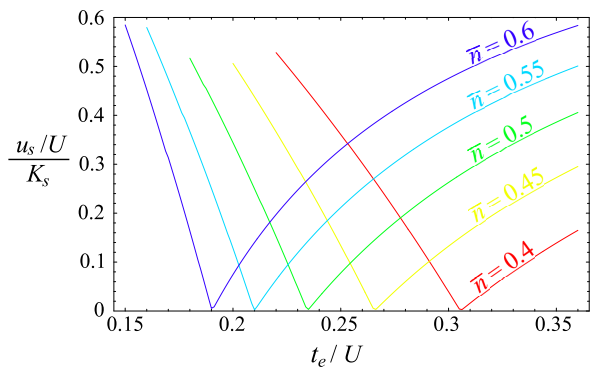}
\caption{The prefactor $u_s/K_s$ of the term $(\partial_x \phi_s)^2$ in the bosonized Hamiltonian near the Stoner point, as a function of $t_e/U$ for varying fillings near half-filling. The red curve stands for $\bar n=0.4$ and the blue line stands for $\bar n=0.6$.}
\label{fig_uOverK_various_fillings}
\end{figure}

Fig. \ref{fig_uOverK_various_fillings} shows that the previous considerations are not special to the magic filling: for all fillings of the edge state, $u_s/K_s$ is non-negative on both sides of the Stoner point, if the proper mean-field theory is used.

Finally, we consider the backscattering term in the ferromagnetic regime. From Fig. \ref{fig_mag_kf}, one can see that backscattering is not momentum-conserving for $\Delta k\neq0$. However, the momentum mismatch is $4\Delta k$ so that close to the Stoner point, the backscattering term is at least approximately momentum-conserving. A careful derivation of the backscattering term gives
\begin{equation}
H_s^1 = \frac{2g_{1\perp}}{(2\pi\eta)^2} \int \D x \cos \left(\sqrt8 \phi_s(x) + 4\Delta k\, x\right).
\end{equation}
For $\Delta k$ sufficiently large, $H_s^1$ averages to zero because of the fast oscillation with $x$ in the cosine; this is just another way of saying that backscattering is momentum non-conserving and thus not allowed.

For small $\Delta k$ the situation is dramatically different, though. Since the prefactor of the free Hamiltonian is very small near the Stoner point, $H_s^1$ is dominant. Thus, the field $\phi_s(x)$ becomes locked to $-4\Delta k x$. This means that the finite mean-field magnetization around which we have expanded the quantum fluctuations is brought back to zero by the backscattering term, and this violates self-consistency. Just as in the non-magnetic regime, our treatment becomes invalid near the Stoner point.

Deeper in the ferromagnetic regime, for $\tau$ sufficiently large, the free Hamiltonian becomes stronger again so that the zero-magnetization state becomes energetically unfavorable. There, $H_s^1$ is not strong enough to destabilize the ferromagnetic mean-field theory and the ferromagnetic ground state is self-consistent. This is quite what one would expect from quantum fluctuations, namely that they reduce the tendency towards a broken symmetry ground state. The critical point, at which the system polarizes spontaneously, is shifted to higher interaction strengths.

\section{Discussion\label{section_discussion}}
In the preceding sections we have demonstrated that edge magnetism at graphene/graphane interfaces can be understood on the basis of a one-dimensional effective model for the edge states. This effective model comprises a non-magnetic regime for a sufficiently large edge state bandwidth, a weak ferromagnetic regime for intermediate bandwidths and a saturated regime for small bandwidths. The saturated regime corresponds to the usual edge magnetism as it has been studied in Refs. \onlinecite{fujita_mf,jung_macdonald_mf,sasaki_saito_mf,son_dft,pisani_dft,feldner_mb,dutta_mb,hikihara_mb}, for instance. The underlying mechanism is based on the one-dimensional version of the well known Stoner instability. The essential difference to previous works is that we utilize the bandwidth $t_e$ of the edge states in order to obtain a well controlled theory. At graphene/graphane interfaces, $t_e$ is experimentally accessible by means of electrostatic gates, so that the edge magnetism can be turned on and off dynamically in an experiment.

Such experimental control over the magnetic state of graphene edges might prove useful for distinguishing between edge magnetism and other sources like disorder as the underlying mechanism for transport gaps in GNRs. 

The advantage of this control parameter for the theory becomes evident by comparing this work to Ref. \onlinecite{hikihara_mb}, where the linear kinetic energy term is absent. There the exponential overlap between adjacent edge states is used in order to define the (non-interacting) kinetic energy term around which the quantum fluctuations can be treated. It is obviously hard to study the effect of quantum fluctuations directly in the fully polarized regime where the band of one spin species is either completely occupied or completely unoccupied, because it is not possible to bosonize this theory in the usual way. In our effective edge state model, instead, we can tune from the usual Luttinger liquid to the ferromagnetic region in a well controlled manner. In the intermediate regime of weak ferromagnetism (not too close to a certain critical point, though), bosonization works well and we are able to study the effect of quantum fluctuations in a ferromagnetic Luttinger liquid.

It should be noted that ferromagnetic Luttinger liquids have been studied by Bartosch {\it et al.}\cite{bartosch}. There, it has been shown that the ferromagnetism can be stabilized by a large positive third derivative of the dispersion. Edge states in graphene or at graphene/graphane interfaces do not fall into this category, however. In graphene edge states, the ferromagnetism is rather stabilized by means of effective velocity-dependent electron-electron interactions: the larger the distance between the Fermi momenta of the different spin species, the smaller is the effective interaction.

The velocity-dependence of the interaction comes from the momentum-dependence of the localization length of the edge state wave function. Therefore, the same mechanism that gives rise to weak ferromagnetism at graphene edges, might be found also in other systems with similar localization properties of edge states, e.g. in topological insulators where the edge states also become more and more delocalized as they merge into the bulk.

Since we are investigating magnetism in one dimension, a comment about the applicability of the Lieb-Mattis\cite{lieb_mattis_theorem} theorem, which forbids ferromagnetic order in one dimension, is mandatory: this theorem is not applicable to edge magnetism because the assumptions of Lieb and Mattis exclude velocity-dependent interaction potentials.  Indeed this velocity-dependence is essential for the stabilization of the weak ferromagnetism. Nevertheless, the statement of stability is only a statement of {\it local} stability. Within the line of argument of the present work, we cannot exclude the presence of another phase with lower energy than the ferromagnetic phase. More effort is needed for a final answer to this question. However, our effective model provides a convenient framework for further investigations. For instance, it allows numerical quantum many-body simulations with fewer effort than for the full two-dimensional lattice model because all unimportant degrees of freedom have been eliminated already, while the important properties of edges of honeycomb lattices have been condensed into a 'small' effective one-dimensional model.

However, there is a further indication which helps to gain more confidence in the existence of the weak ferromagnetic regime: in the limit $t_e\rightarrow0$, the two-dimensional honeycomb lattice model becomes particle-hole symmetric so that Liebs theorem\cite{liebs_theorem} applies and predicts a high-spin ground state at a zigzag edge. The particle-hole symmetric case corresponds to the completely saturated regime in the present work. This means that our prediction of weak ferromagnetism at graphene/graphane interfaces is consistent with the well accepted limits (a) edge magnetism in graphene for $t_e\rightarrow0$ and (b) Luttinger liquid behavior for $U\ll |t_e|$.

Another important issue, connected to the low dimensionality, is the question of the stability of edge magnetism w.r.t. finite temperatures. Of course, the usual entropy argument, which prohibits a spin-polarization in a one-dimensional system in the thermodynamic limit, is applicable here. However, graphene structures are usually of mesoscopic size, which means that spin-waves exhibit a finite-size energy gap. Thus, the ground state, which we have studied in this work, should be observable for sufficiently small mesoscopic structures at sufficiently low temperatures.

\begin{acknowledgments}
We acknowledge useful discussions with B. Braunecker. This work has been supported by the Swiss NF and the NCCR Nanoscience Basel.
\end{acknowledgments}

\appendix

\section{Graphane gate\label{appendix_graphane_gate}}

An atomically sharp gate as described above is experimentally not feasible. Even if the termination of such a metallic gate could be controlled on an atomic scale, the potential would leak into the bulk graphene region because of the weak screening in graphene, thus changing the electron density in the graphene region.

To circumvent this, we propose to use gates of different voltage on different sides of the plane. This gate lifts (lowers) the on-site energy of the hydrogen atoms on the $+z$ ($-z$) side of the $z=0$ plane. Since the graphene atoms are located at $z=0$, the on-site energy is unchanged in the graphene region. We use the effective model of edge states at graphene/graphane interfaces from Ref. \onlinecite{our_EA_paper} which is defined by the Hamiltonian
\begin{equation}
H_{\rm eff} = \begin{pmatrix} 0 & t\mathcal N_k & 0 \\ t\mathcal N_k & 0 &t'\\0&t'&\epsilon_H\end{pmatrix},
\end{equation}
where $t\simeq -3$ eV is the usual hopping between carbon $\pi$ orbitals, $t'\simeq -6$ eV is the hopping between the carbon $\pi$ orbital and the hydrogen 1s orbital in the graphane region, and $\epsilon_H \simeq -0.4$ eV is the hydrogen 1s orbital energy without any gates. Since $\epsilon_H$ is smaller than all other energy scales, we treat it in perturbation theory. Setting $\epsilon_H=0$, we obtain a zero energy eigenstate
\begin{widetext}
\begin{equation}
\left|\psi(k)\right>= -\left[\left(\frac{\mathcal N_k^\alpha t}{t'}\right)^2+1\right]^{-\frac12} \left|\psi_0^\alpha(k)\right> + \left[1+\left(\frac{t'}{\mathcal N_k^\alpha t}\right)^2\right]^{-\frac12} \left|H\right> \simeq \left[1-\frac12 \left(\frac{\mathcal N_k^\alpha t}{t'}\right)^2\right]\left|\psi_0^\alpha(k)\right> + \mathcal N_k^\alpha\frac{ t}{t'} \left|H\right>. \label{interface_state_wf}
\end{equation}
\end{widetext}
From Eq. (\ref{interface_state_wf}), we see that the wave function of the edge state at a graphene/graphane interface lives predominantly at the graphene region sites where the usual graphene edge state lives ($\left|\psi_0(k)\right>$) and at the hydrogen atom in the first graphane row. Obviously then, a non-zero on-site energy at the hydrogen atom leads in first order perturbation theory to an energy shift of the order of this on-site energy. The hydrogen on-site energy is composed of the intrinsic chemical potential of hydrogen $\epsilon_H $ and the gate-induced on-site potential $\Delta$. Thus, we find
\begin{equation}
\epsilon^\alpha(k) = (\epsilon_H + \Delta) \frac{t^2}{t'^2}(2\cos(k-\pi) -1).
\end{equation}

The distance between the hydrogen atom planes 2.8\AA\, and the distance between the carbon planes is 0.6\AA. Thus, if the hydrogen atom is at the potential $\Delta$ then the carbon atom attached to this hydrogen atom is roughly at the potential $0.2\Delta$. This gives an additional positive contribution to the edge state bandwidth $(\epsilon_H + \Delta)t^2/t'^2$, since the edge state wave function has also a weight of order $\epsilon_H/t$ on this carbon atom.

The largest electric fields that can be reached are of order $10^7-10^{8}$ V/cm. Thus, the total hydrogen on-site potential can be of the order of one eV, leading to an in-situ tunable bandwidth range of a several hundred meV in addition to the edge state bandwidth contributions coming from, e.g., farther neighbor hoppings. Also, the bandwidth can be tuned by substituting the hydrogen atoms in the graphane region with other elements or molecules with different orbital energies.

\section{Effective Hubbard interaction\label{appendix_eff_hubbard}}

The edge state operator reads
\begin{equation}
e_{k,\sigma} = \mathcal N_k \sum_{n=0}^\infty [-u_k^*]^n d_{n,k,B,\sigma}
\end{equation}
(we drop the sublattice index henceforth because the edge state lives on the B sublattice only) and its commutation relation with the $d$-operator is
\begin{equation}
\left\{e_{k\sigma},d^\dagger_{n,k,\sigma} \right\} = \mathcal N_k [-u^*_k]^n \delta_{kk'}\delta_{\sigma\sigma'}.
\end{equation}
Now we project the Hubbard Hamiltonian (restricted to B sites)
\begin{equation}
H_U = UN_x^{-1} \sum_{k,k',q,n} d^\dagger_{k+q,n,\uparrow} d_{k,n,\uparrow} d^\dagger_{k'-q,n,\downarrow} d_{k',n,\downarrow}
\end{equation}
to the Fock space spanned by the edge states ($e_{i\sigma}$ is a short form for $e_{k_i,\sigma}$)
\begin{equation}
\tilde H_U =  \sum_{k_1,k_2,k_3,k_4}' \left<1\uparrow,2\downarrow | H_U | 3\uparrow,4\downarrow\right> e_{1\uparrow}^\dagger e_{2\downarrow}^\dagger e_{4\downarrow} e_{3 \uparrow},
\end{equation}
where the two-fermion states are $\left|1\sigma,2\sigma'\right>=e_{1\sigma}^\dagger e_{2\sigma'}^\dagger\left|0\right>$. The primed sum is restricted to values of $\frac{2\pi}3<k_i<\frac{4\pi}3$. We calculate the matrix element
\begin{equation}
\frac{U}{N_x} \sum_{k,k',q,n} \left<0\right| e_{2\downarrow} e_{1\uparrow} d^\dagger_{k+q,n,\uparrow} d_{k,n,\uparrow} d^\dagger_{k'-q,n,\downarrow} d_{k',n,\downarrow} e_{3\uparrow}^\dagger e_{4\downarrow}^\dagger \left|0\right>
\end{equation}
and find
\begin{equation}
\tilde H_U = \frac{U}{N_x} \sum_{k,k',q}' \Gamma(k,k',q) e_{k+q,\uparrow}^\dagger e_{k,\uparrow} e_{k'-q,\downarrow}^\dagger e_{k',\downarrow} 
\end{equation}
where the primed sum is restricted to values of $k,k',k+q,k'-q$ to the domain of the $\alpha$-type edge state and
\begin{equation}
\Gamma(k,k',q) =  \frac{\mathcal N^\alpha_{k+q} \mathcal N^\alpha_{k} \mathcal N^\alpha_{k'-q} \mathcal N^\alpha_{k'}}{1-u_{k+q}^* u_k u^*_{k'-q} u_{k'}}.
\end{equation}

\section{Background density at an edge\label{appendix_background_density}}
For convenience, we neglect the spin-degree of freedom in the following discussion. Since we do not use explicitly spin-dependent Hamiltonians, like the spin-orbit interaction, the electron spin only leads to a factor of 2. The strategy of the following argumentation is to show that a half-filled ground state of an $\alpha\alpha$-ribbon has an electron density of exactly $\frac12$ (or 1 with spin) at each site, including the sites at the edges. Then, it is shown that at half-filling, half of the edge states are occupied. Subtracting the part of the electron density, originating from those filled edge states, from the uniform density leads then to the electron density of a state in which all valence (conduction) band states are filled (empty) and all edge states are empty.

We consider an $\alpha\alpha$-ribbon with $N_y$ unit cells in the transverse direction, described by the Hamiltonian
\begin{equation}
H = \sum_{n=1}^{N_y-1} d^\dagger_{n,k,A}d_{n,k,B} + \sum_{n=1}^{N_y} u_k d^\dagger_{n,k,A} d_{n-1,k,B} + h.c.
\end{equation}
and an odd number $N_x$ of unit cells in $x$-direction, in order to exclude $k=\pi$. Because of this exclusion, there are no states with exactly zero energy. Only eigenenergies which are exponentially small in $N_y$ exist. The particle-hole transformation $U=U^\dagger$ acts onto the $d$-operators
\begin{equation}
d_{n,k,A}\rightarrow d_{n,k,A},\bs\bs d_{n,k,B}\rightarrow -d_{n,k,B}.
\end{equation}
Thus, $U H U^\dagger = -H$. The density operators $d^\dagger_{n,k,s}d_{n,k,s}$ are invariant under $U$. We now fill exactly half of the states
\begin{equation}
\left|\chi_0\right> = \prod_n \Theta(-\epsilon_n) a_n^\dagger \left|0\right>
\end{equation}
where $a_n^\dagger$ creates the eigenstate $n$ with eigenvalue $\epsilon_n$. The dual state is $\left|\bar\chi_0\right> = U\left|\chi_0\right> = \prod_n \Theta(\epsilon_n) a_n^\dagger \left|0\right>$. Because of the particle-hole antisymmetry of the Hamiltonian and particle-hole symmetry of the density operators we have
\begin{equation}
\left<\chi_0\right|d^\dagger_{n,k,s}d_{n,k,s}\left|\chi_0\right> = \left<\bar\chi_0\right|d^\dagger_{n,k,s}d_{n,k,s}\left|\bar\chi_0\right> =\frac12.
\end{equation}
The states of which $\left|\chi_0\right>$ is composed are all valence band states plus only the antisymmetric (or only the symmetric) combinations of the oppositing edge states. This is because $U$ turns antisymmetric edge state combinations into symmetric combinations. Thus, the part of the electron density, $\frac12$ per site, deriving from the edge state wave functions $\psi_{0,k}(n,s)$ is
\begin{equation}
\frac12 \int_{\frac{2\pi}3}^{\frac{4\pi}3}\frac{\D k}{2\pi} |\psi_{0,k}(n,s)|^2 = \frac12 \rho_0(n,s),
\end{equation}
where $\rho_0(n,s)$ has been defined in Eq. (\ref{half_filled_edge_density}). It follows that the background density is
\begin{equation}
\tilde\rho_B(n,s) = \frac12 - \frac12 \rho_0(n,s)
\end{equation}
relative to zero filling, or
\begin{equation}
\rho_B(n,s) = \tilde\rho_B(n,s)-\frac12 = -\frac12 \rho_0(n,s)
\end{equation}
relative to half-filling.

\section{Comparison to a numerical calculation\label{appendix_numerics}}
In the main part of this paper, we treated the interaction as a perturbation to the hopping Hamiltonian $H_0$. Now, we perform numerical lattice-based mean-field calculations for finite size ribbons. Our numerical calculation for an $\alpha\alpha$-ribbon with $N$ unit cells in the transverse direction is based on the Hamiltonian
\begin{multline}
H = t\sum_{k,n,\sigma} d^\dagger_{n,k,A,\sigma}(d_{n,k,B,\sigma}+u_k d_{n-1,k,B,\sigma}) + h.c. \\
- t_e \sum_{k,\sigma} \left[d^\dagger_{0,k,B,\sigma} d_{0,k,B,\sigma} + d^\dagger_{N,k,A,\sigma} d_{N,k,A,\sigma} \right]\\
+\frac{U}{N_x} \sum_{n,q,s} \left[\hat\rho_{n,q,s,\uparrow} \left<\hat\rho_{n,-q,s,\downarrow}\right> + \left<\hat\rho_{n,q,s,\uparrow} \right>\hat\rho_{n,-q,s,\downarrow}  \right],\label{numerical_hamiltonian}
\end{multline}
with the spin-dependent densities
\begin{equation}
\left<\hat\rho_{n,q,s,\sigma}\right> = \delta_{q,0} \int \frac{\D k}{2\pi} |\psi_{m,k,\sigma}(n,s)|^2 \Theta\left[\epsilon_F - \epsilon_{m,\sigma}(k)\right],\label{app_density}
\end{equation}
where $\psi_{m,k}(n,s,\sigma)=\left<0\right|d_{n,k,s,\sigma}\left|m,k,\sigma\right>$ is the wave function of the eigenstate to the energy $\epsilon_{m,\sigma}(k)$. $t = -3$eV is the hopping amplitude for nearest neighbor hopping of graphenes $\pi$-band, $t_e$ is the strength of the edge gate by which we model a certain class of edge/interface properties (see Sect. II), and $U$ is the strength of the Hubbard interaction.

We choose a discrete set of about 4000 k points between 0 and 2$\pi$ in order to approximate the integral in Eq. (\ref{app_density}). The Fermi energy is chosen such that exactly half of all eigenstates are filled. We start with a density that has a small positive magnetization on the one edge and a small negative magnetization at the other edge, i.e. $\left<\hat\rho_{n,0,s,\sigma}\right>_{\rm init} = \frac12 + \frac\sigma{10}\left(\delta_{n,0} \delta_{s,B} - \delta_{n,N}\delta_{s,A}\right)$, calculate the eigenvalue and eigenmodes of (\ref{numerical_hamiltonian}) and from it a new set of spin-dependent densities by Eq. (\ref{app_density}). This procedure is then iterated until the densities do not change anymore, i.e. self-consistence is reached.

\begin{figure}[!ht]
\centering
\includegraphics[width=210pt]{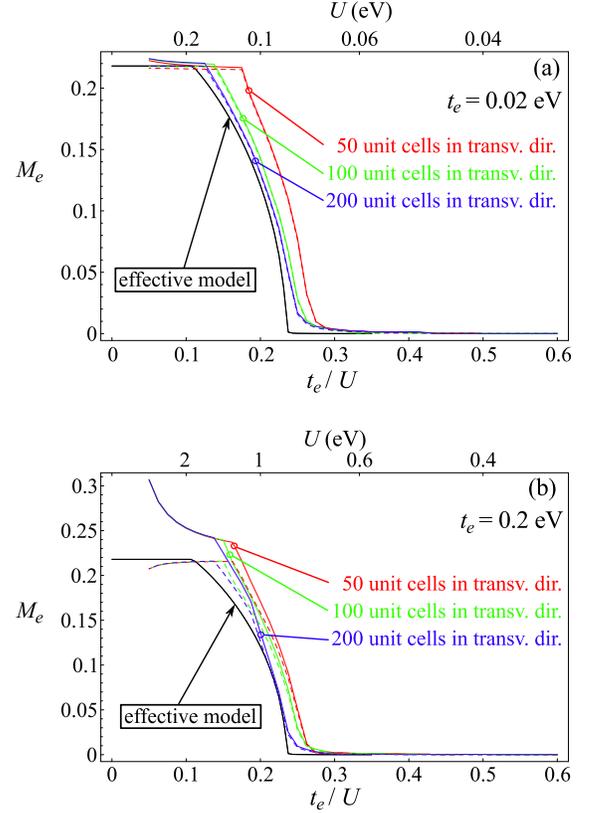}
\caption{(Color online) Comparison of the edge magnetizations extracted from the result of the numerical calculation and from the effective model. The solid black line is the edge magnetization from the effective model. The solid gray lines show $M_e^{\rm num}$ of an $\alpha\alpha$-ribbon with different widths (50, 100, and 200 unit cells in the transverse direction). The lower abscissas of both parts of the figure shows $t_e/U$ where the band width parameter has been fixed to $t_e=0.02$eV in Part (a) and $t_e=0.2$eV in Part (b). The corresponding Hubbard interaction strength $U$ is shown in the upper abscissas. The dashed lines show the polarization of the edge states $M_e^{\rm num,es}$.}
\label{fig_comp_ana_num}
\end{figure}

As a quantifier of the solution we choose the edge magnetization. We calculate this quantity for one edge from the self-consistent result of the numerical calculation, i.e.
\begin{equation}
M_e^{\rm num} = \sum_\sigma \sigma \left<\rho_{n=0,q=0,B,\sigma}\right>_{s.c.}.
\end{equation}
Note that the $M_e^{\rm num}$ is not equal to $M_e$, as defined in Eq. (\ref{edge_magnetization}). The edge magnetization $M_e$ of the effective model only respects the spin-polarization of the edge states. $M_e^{\rm num}$, on the other hand, also takes into account the polarization of the bulk states near the edges. This additional spin-polarization from the bulk states is small for sufficiently weak $U$ (see Fig. \ref{fig_comp_ana_num}(a)), while for large $U$, the edge magnetization is considerably enhanced (see Fig. \ref{fig_comp_ana_num}(b)). However, this does not mean that the effective edge state model would be insufficient for describing the edge magnetism. It only means that the polarization of the edge states induces an additional polarization in the bulk states via the strong Hubbard interaction which increases $M_e^{\rm num}$. In this regime, the edge states are already completely polarized. Thus, the difference between the effective model and the numerical calculation is only quantitative, as desired.

Further confidence in the validity of the effective model can be gained by  directly calculating the polarization of the edge states, i.e.
\begin{equation}
M_e^{\rm num,\, es} = \sum_\sigma \sigma \int_{\frac{2\pi}3}^{\frac{4\pi}3} \frac{\D k}{2\pi} |\psi_{0,k,\sigma}(0,B)|^2 \Theta\left[\epsilon_F - \epsilon_{0,\sigma}(k)\right],
\end{equation}
where $\psi_{0,k,\sigma}(n,s)$ is the wave function of the edge state with spin $\sigma$ in unit cell $n$ and sublattice site $s$, with energy $\epsilon_{0,\sigma}(k)$.

$M_e^{\rm num,\,es}$ is plotted, for different ribbon widths and $t_e$, as dashed lines in Fig. \ref{fig_comp_ana_num}. Compared to $M_e^{\rm num}$, it is much closer to the result of the effective model, as expected. Nevertheless, the saturation polarization of the edge states is reached already for smaller $U$. This can be interpreted as a back action of the induced polarization of the bulk states on the edge states which acts like an additional magnetic field. Note that this interpretation is in consistence with Ref. \onlinecite{fRG_jutta}, where it is found that integrating out the bulk states in graphene structures leads to enhanced effective interaction parameters for the edge states. As expected, we observe that such mechanisms are only important for large $U$ (see Fig. \ref{fig_comp_ana_num}(b)) while they are absent for small $U$ as can be seen from Part (a) of Fig. \ref{fig_comp_ana_num}.

Thus, the following physical picture emerges from the comparison between the numerics and the analytical model: the edge states are primarily responsible for the edge magnetization. The polarization of the edge states then induces an additional spin-polarization in the bulk states if the Hubbard interaction is large enough. This additional bulk state polarization further increases the edge state polarization so that the saturation is reached already for smaller $U$. The essential approximation, we have made in the effective edge state model, is that we neglected the enhancement of the effective interaction by the bulk states. This enhancement can be easily reintroduced into the model, if desired.

\end{document}